\documentclass[aps,prl,twocolumn,superscriptaddress]{revtex4-1}

\usepackage[latin9]{inputenc}
\usepackage{color}
\usepackage[colorlinks,urlcolor=blue,citecolor=blue,linkcolor=blue]{hyperref}
\usepackage{verbatim}
\usepackage{float}
\usepackage{amsmath}
\usepackage{amssymb}
\usepackage{graphicx}
\usepackage{epstopdf}
\usepackage{esint}
\usepackage{CJK}
\usepackage{braket}
\definecolor{olive}{RGB}{0,100,0}

\begin{document}
\begin{CJK*}{UTF8}{gbsn} 
\title{Breathing Mode of a BEC Repulsively Interacting with a Fermionic Reservoir}

\author{Bo Huang (黄博)}
\email{Bo.Huang@uibk.ac.at}
\affiliation{Institut f\"ur Quantenoptik und Quanteninformation (IQOQI), \"Osterreichische Akademie der Wissenschaften, 6020 Innsbruck, Austria}
\author{Isabella Fritsche}
\thanks{I.F. and B.H. contributed equally to this work.}
\author{Rianne S. Lous}
\author{Cosetta~Baroni}
\affiliation{Institut f\"ur Quantenoptik und Quanteninformation (IQOQI), \"Osterreichische Akademie der Wissenschaften, 6020 Innsbruck, Austria}
\affiliation{Institut f\"ur Experimentalphysik und Zentrum f\"ur Quantenphysik, Universit\"at Innsbruck, 6020 Innsbruck, Austria}

\author{Jook T. M. Walraven}
\affiliation{Institut f\"ur Quantenoptik und Quanteninformation (IQOQI), \"Osterreichische Akademie der Wissenschaften, 6020 Innsbruck, Austria}
\affiliation{Van der Waals-Zeeman Institute, Institute of Physics, University of Amsterdam,
Science Park 904, 1098 XH Amsterdam, The Netherlands}

\author{Emil Kirilov}
\affiliation{Institut f\"ur Experimentalphysik und Zentrum f\"ur Quantenphysik, Universit\"at Innsbruck, 6020 Innsbruck, Austria}
\author{Rudolf Grimm}
\affiliation{Institut f\"ur Quantenoptik und Quanteninformation (IQOQI), \"Osterreichische Akademie der Wissenschaften, 6020 Innsbruck, Austria}
\affiliation{Institut f\"ur Experimentalphysik und Zentrum f\"ur Quantenphysik, Universit\"at Innsbruck, 6020 Innsbruck, Austria}
\date{\today}
\begin{abstract}
We investigate the fundamental breathing mode of a small-sized elongated Bose-Einstein condensate coupled to a large Fermi sea of atoms. 
Our observations show a dramatic shift of the breathing frequency when the mixture undergoes phase separation at strong interspecies repulsion. We find that the maximum frequency shift in the full phase-separation limit depends essentially on the atom number ratio of the components. We interpret the experimental observations within a model that assumes an adiabatic response of the Fermi sea, or within another model that considers single fermion trajectories for a fully phase-separated mixture. These two complementary models capture the observed features over the full range of interest.
\end{abstract}


\maketitle
\end{CJK*}


Mixtures of quantum fluids play a fascinating role in our understanding of multi-component many-body quantum systems. For decades, the study of such mixtures focused on the phases of the helium isotopes $^3$He and $^4$He and their properties in mixed states, under phase-separated conditions, or at the interface between two phases~\cite{Ebner1971tlt}. Ultracold atomic gases have opened up many new opportunities, and various weakly and strongly interacting Bose-Bose, Fermi-Fermi and Bose-Fermi mixtures have been investigated \cite{Pitaevskii2016book, Pethick2002book}. A unique feature of ultracold quantum gases is the possibility to tune the interparticle interactions over a wide range by magnetically controlled Feshbach resonances (FRs)~\cite{Chin2010fri}. 

Right from the early experiments on harmonically trapped quantum gases \cite{Jin1996ceo, Mewes1996ceo}, collective modes have served as powerful probes for interparticle interactions. Depending on their particular character \cite{Pitaevskii2003book}, collective modes can be sensitive to different effects. If the trapped sample changes its position, angle, or form without undergoing significant volume changes, the mode can be classified as a surface mode. Excitations of this kind have been used to study the transition from hydrodynamic to collisionless behavior in both bosonic \cite{StamperKurn1998cah,Buggle2005soi} and fermionic~\cite{Altmeyer2007doa, Wright2007ftc} quantum gases. If, in contrast, the oscillation involves significant changes of the volume and thus of the density of the sample, then the mode can be classified as a compression or breathing mode. Modes with predominant compression character can serve as sensitive probes for the equation of state. As an example, the radial breathing mode in an elongated trap~\cite{Stringari1996ceo,Chevy2002tbm} has served as a tool to probe strongly interacting Fermi gases~\cite{Kinast2004efs, Bartenstein2004ceo, Altmeyer2007pmo}.

In ultracold atomic mixtures, the motional coupling generally leads to rich behavior in the collective modes, see e.g.\ Refs.~\cite{Busch1997sac, Esry1998lle, Ho1998sbc, Bijlsma2000pei, Yip2001cmi, Capuzzi2001zsd, Pu2002psa, Liu2003cah, Rodriguez2004smo} for early theoretical considerations. As a basic example, the center-of-mass oscillations of different components (their so-called dipole modes), which can experience frequency shifts and damping \cite{Vichi1999coo, Maddaloni2000coo, Gensemer2001tfc}, have been utilized in recent experiments to study coupling effects in mixed superfluids \cite{FerrierBarbut2014amo, Delehaye2015cva, Roy2017tem, Wu2018cdo} and to investigate mediated interactions \cite{DeSalvo2018fmi}. In the case of strong interspecies interactions, instabilities (collapse \cite{Modugno2002coa,Ospelkaus2006idd} or phase separation \cite{Zaccanti2006cot,Ospelkaus2006toh, Papp2008tmi,Shin2008pdo,Valtolina2017etf, Lous2018pti}) render the collective dynamics even more complex. Although the understanding of collective behavior near instabilities is essential in view of proposed fermionic superfluids based on mediated pairing \cite{Bijlsma2000pei,Heiselberg2000ioi, Efremov2002pwc,Suzuki2008pws, Enss2009snp,Caracanhas2017fbm, Kinnunen2018ipw}, corresponding experimental information is rather scarce.

\begin{figure}[t]
\includegraphics[width=0.4\textwidth]{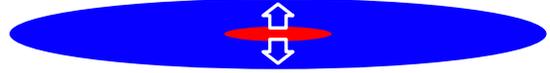}
\caption{Radial breathing mode of a small BEC (red) residing in a large Fermi sea (blue). The atomic quantum-gas mixture is kept in a highly elongated optical trap.}
\label{fig:breathingcigar}
\end{figure}

In this Letter, we consider the fundamental breathing mode of a Bose-Einstein condensate (BEC) repulsively coupled to a large fermionic reservoir of atoms. Our system, realized with optically trapped $^{41}$K bosons and $^6$Li fermions (see Fig.~\ref{fig:breathingcigar}), offers tunable interspecies interaction and allows us to explore the regime of strong repulsion, where the BEC phase separates from the surrounding fermions \cite{Lous2018pti}. As a dynamic phenomenon resulting from phase separation, we demonstrate the emergence of a drastic up-shift of the BEC's breathing mode frequency. We show how this effect depends on the interaction strength and on the atom number ratio of bosons and fermions. We interpret the complex dynamical many-body physics of our system in terms of two complementary models, which capture the observed features over the full range of interest.

The frequency shift can be understood qualitatively by considering the interface that emerges from phase separation of the Bose-Fermi mixture. In the presence of the interface, the BEC becomes hydrostatically compressed by the Fermi pressure. Exciting a collective mode of the BEC leads to a motion of this interface. If the mode is a breathing mode, the oscillation inflates and deflates the interface, like modulating a bubble in the Fermi sea. Intuitively, the volume change of the BEC leads to a significant reversible work against the Fermi pressure. Because of the existence of this strong restoring mechanism the oscillation frequency substantially increases. This stands in contrast to surface modes of BECs immersed in Fermi gases, which have been observed in  experiments~\cite{FerrierBarbut2014amo, Delehaye2015cva,  Roy2017tem, DeSalvo2018fmi, Wu2018cdo}. There, the frequency shifts are rather small, since surface modes do not change the volume and thus do no work against the Fermi pressure.

The transition into the phase-separated regime is characterized by two distinct values of the interspecies scattering length $a_{bf}$. We define a depletion scattering length $a_d$ as the value at which the fermion density drops to zero in the center of the trap, where one finds the highest boson density. The value of $a_d$ is trap specific and depends on the densities of the components, so we obtain it numerically~\cite{Lous2018pti}. We also define a critical scattering length 
\begin{equation}\label{eq:ac}
a_c=\frac{\sqrt{15\pi}}{4} r_m \sqrt{\frac{a_{bb}}{k_{\rm F}}}
\end{equation}
as the value where the mixture fully phase separates in the Thomas-Fermi limit \cite{Viverit2000ztp}.
Here $r_m=2\sqrt{m_bm_f}/(m_b+m_f)$, $m_b$ and $m_f$ are the boson and the fermion masses respectively,  $k_{\rm F}=\sqrt{2m_f E_{\rm F}/\hbar^2}$ is the Fermi wave number, and $a_{bb}$ is the boson-boson scattering length. We note that, under the realistic experimental conditions of a system of finite size, the phase transition is smoothed by the kinetic energy of the BEC \cite{Lous2018pti}.

Our $^{41}$K-$^6$Li mixture is produced via laser and evaporative cooling following a procedure described in Ref.~\cite{Lous2018pti} and kept in an elongated optical dipole trap, which is formed by two crossed infrared laser beams and has an aspect ratio of 7.6~\cite{ODT2beams}.
 The radial trap frequency is  $\omega_{b}=2\pi\times 171$ Hz for K and $\omega_{f}=2\pi\times 300$ Hz for Li. Typically, we have a sample of $10^5$ Li atoms in the lowest spin state Li$|1\rangle$ ($F=1/2$, $m_F=1/2$) and $4\times 10^4$ K atoms prepared in the second-to-lowest spin state K$|2\rangle$ ($F=1$, $m_F=0$). The mixture is thermalized at a temperature of $T/T_{\rm F}\approx 0.13$, where $T_{\rm F}\approx 700$ nK is the Fermi temperature of the Li cloud. With a condensate fraction of about 1/3, the K BEC has an atom number of $\sim 2 \times 10^4$. 

We vary the interspecies interaction strength by a combination of spin-state manipulation and Feshbach tuning. First, we control the particular spin state of the K atoms by application of radio-frequency (rf) $\pi$-pulses. In the case of the Zeeman sublevel K$\ket{1}$ ($F=1$, $m_F=1$), a Feshbach resonance near 335G facilitates tuning of the interspecies scattering length according to $a_{bf}=a_{\rm bg}[1-\Delta/(B-B_0)]$, where $a_{\rm bg}=60.9a_0$ ($a_0$ is Bohr's radius), $\Delta=0.949$ G and $B_0=335.057(1)$ G~\cite{lightshift}.  In the case of K$\ket{2}$, only the weak background interaction is present ($a_{bf}\approx 60a_0$),  which provides enough thermalization for sympathetic cooling between the two species, but is too weak to induce significant changes to the density profiles~\cite{Lous2018pti}. The boson-boson scattering length stays constant  as $a_{bb}=60.9a_0$~\cite{abb}.

\begin{figure}[t]
\includegraphics[width=0.4\textwidth]{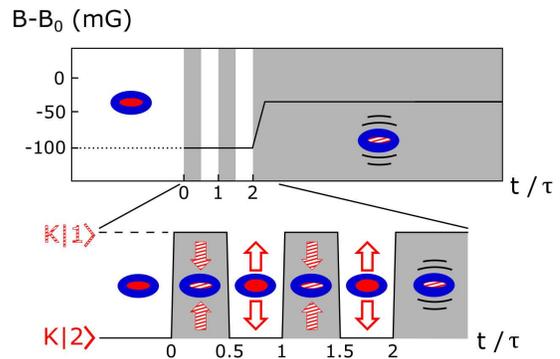}
\caption{Excitation procedure for the breathing oscillation of the BEC with a series of interaction quenches. The magnetic field is first set to a value where $a_{bf}\approx 700a_0$  in the K$\ket{1}$-Li$\ket{1}$ mixture, but we start in the K$\ket{2}$-Li$\ket{1}$ mixture, where the interaction is weak ($a_{bf}\approx 60a_0$). Then we apply several radio-frequency $\pi$-pulses where the time between consecutive flips is $\tau/2$, which corresponds to half of the radial breathing mode period of the BEC in the absence of fermions. After this multiple-pulse excitation we ramp the $B$-field to the target value and let the oscillation continue there.}
\label{fig:excitation}
\end{figure}

To excite the breathing mode of the K condensate we modulate the interspecies interaction
by periodically changing the scattering length~\cite{Matthews1998dro,Pollack2010ceo} . As illustrated in Fig.~\ref{fig:excitation}, this is achieved by alternating
the state of the K atoms between K$\ket{2}$ ($a_{bf}\approx 60a_0$) and K$\ket{1}$ ($a_{bf}\approx 700a_0$) using a short burst
of rf-pulses at $B-B_0$=$-100$ mG. The $\pi$-pulse duration is 100 $\mu$s, which is much shorter than the pulse
spacing of 1.4 ms. The latter is roughly matched to half the period of the radial breathing mode
(full period $\tau\approx\pi/\omega_b$) in order to resonantly drive the oscillation. Starting in K$\ket{2}$, a burst of three
rf pulses enables us to excite the  fundamental breathing mode of the K$\ket{1}$ condensate (which is mostly radial) with a $\pm$25\% modulation
of the radial size, accompanied by a much slower oscillation in the axial size \cite{SM}. The duration of the
burst affects the oscillation amplitude but has no noticeable influence on the measured breathing
mode frequency $\omega$. Within our detection limits, we do not observe oscillations in the thermal cloud
of K atoms or in the Li cloud.

Immediately after the excitation stage, the $B$-field is ramped within 1~ms to the target value of $a_{bf}$. We hold the excited mixture for a variable time and then we switch off the trap and take time-of-flight images of the expanding atomic clouds. To obtain the frequency $\omega$ of the breathing mode, we fit the recorded time evolution of the width of the BEC with a damped harmonic oscillation with a slowly varying background, the latter being caused by the small residual excitation of the axial mode~\cite{SM}. Typically we record about six breathing mode periods for each measurement, as a longer hold time can only marginally  improve the precision of the measurement~\cite{SM}. 
Moreover, we can safely ignore the influence of atom number decay on $\omega$ within this short period, since the recombination loss from the BEC is below 20\% at most of the values of $a_{bf}$ and still smaller than 50\% for the few points taken very close to the FR center ($a_{bf}>2000a_0$). 

In order to normalize $\omega$ we measure corresponding value $\omega_0$ in the limit of small $a_{bf}$. This is accomplished by adding an additional $\pi$-pulse to the above excitation sequence to prepares a K$\ket{2}$ BEC, which provides a good approximation of the non-interaction case. We verified the expected relation to the trap frequency $\omega_0=2\omega_{b}$ for the radial breathing mode of an elongated BEC~\cite{Stringari1996ceo,Chevy2002tbm}  within a 1$\sigma$ uncertainty of 2\%.

\begin{figure}[b]
\includegraphics[width=0.5\textwidth]{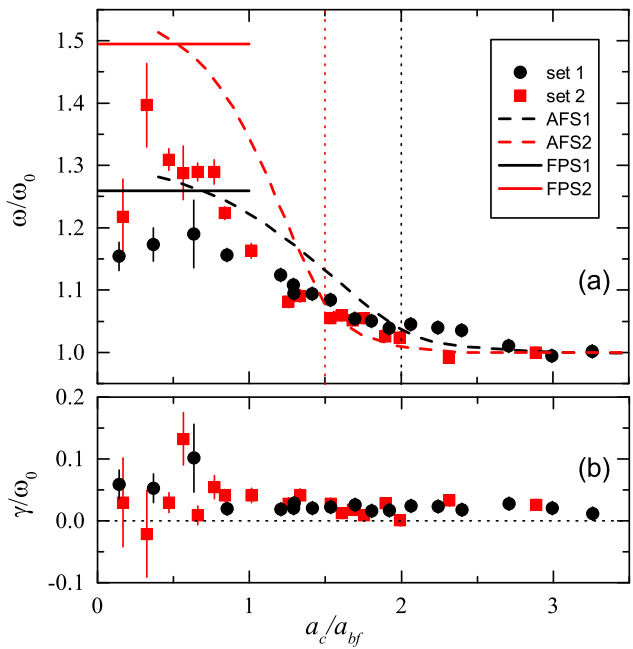}
\caption{BEC breathing mode frequency $\omega$ and damping rate $\gamma$ as a function of the dimensionless interaction parameter $a_c/a_{bf}$. 
The two sets of measurements (filled black circles and red squares) correspond to different atom number ratios (see text). Both observables are normalized to $\omega_0$. The vertical dotted lines indicate the corresponding $a_d/a_{bf}$. The theoretical results from the AFS and FPS model are plotted as dashed and solid curves with corresponding color for the two sets of measurements. The error bars indicate the 1$\sigma$ fitting uncertainty.}\label{fig:vary_abf}
\end{figure}

In Fig.~\ref{fig:vary_abf}(a) we present our measurements of $\omega/\omega_0$ as a function of the dimensionless interaction parameter $a_c/a_{bf}$ with $a_c$ according to Eq.~\eqref{eq:ac}. The first set of measurements (filled black circles) was taken with the boson number $N_b=1.6\times 10^4$ and the fermion number $N_f=1.0\times 10^5$ ($N_b/N_f=0.16$), for which we calculate the two characteristic values of the scattering length as $a_c=619a_0$ and $a_d=308a_0$ ($a_c/a_d=2.0$). For increasing repulsive strength, i.e.~decreasing $a_c/a_{bf}$, we first observe a slow increase of $\omega/\omega_0$ until $a_c/a_{bf} \approx 2$ is reached. Here the fermion cloud becomes fully depleted in the center of the BEC \cite{Lous2018pti}. Then, in the intermediate range of $a_c/a_{bf}$ between 2 and 1, $\omega/\omega_0$ rapidly rises until a plateau value of about 1.2 is reached. For even stronger repulsion in the phase-separated regime, no further frequency change is observed. 
 These results show that the frequency up-shift emerges exactly where the transition to the phase-separated regime occurs and finally levels off at the plateau value when full phase separation is reached. 

We conducted a second set of measurements with a different atom number ratio ($N_b=8.0\times 10^3$, $N_f=1.7\times 10^5$ and thus $N_b/N_f=0.05$)~\cite{SM}, for which $a_c=595a_0$ and $a_d=400a_0$ ($a_c/a_d=1.5$). The corresponding results (set 2) are presented as filled red squares in Fig.~\ref{fig:vary_abf}. In comparison with set 1, the breathing mode frequency in set 2 starts to increase at a slightly smaller value of $a_c/a_{bf}$. But the increase of $\omega/\omega_0$ is steeper, and a higher plateau value around 1.3 is reached. This is qualitatively expected since a smaller value of $N_b/N_f$ corresponds to a more strongly compressed BEC and thus a larger frequency change.

For completeness we also show the normalized damping rate $\gamma/\omega_0$ as a function of the interaction parameter $a_c/a_{bf}$ in Fig.~\ref{fig:vary_abf}(b). In the region of $a_c/a_{bf}>1$ we observe a nearly constant value of $\gamma/\omega_0\approx  0.02$, and we attribute this weak damping to the anharmonicity of the crossed optical dipole trap and the interaction between the BEC and the non-condensate bosons~\cite{Jin1996ceo,damping}.   In the phase-separated regime where $a_c/a_{bf}<1$, the damping rate shows a trend towards higher values with larger uncertainties, 
which may be due to a residual excitation of higher-order radial modes.

\begin{figure}[b]
\includegraphics[width=0.5\textwidth]{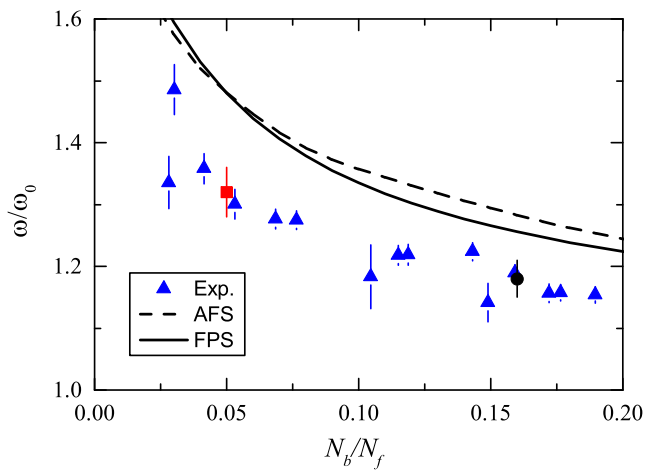}
\caption{Breathing mode frequency in the phase-separation limit as a function of $N_b/N_f$. The blue triangles show the measurements performed at a fixed $a_{bf}=1330a_0$ ($a_c/a_{bf}\approx 0.45$). The error bars show the 1$\sigma$ fitting error. The black circle and the red square show the frequency shift at $a_{bf}\approx1330a_0$ as extracted from sets 1 and 2 in Fig.~\ref{fig:vary_abf}. The numerical curves from the AFS (dashed) and FPS (solid) model are calculated for a total atom number of $1.5\times10^5$~\cite{SM}.}\label{fig:vary_Q}
\end{figure}

Motivated by the observed different plateau values of $\omega/\omega_0$ in the phase-separated regime, we further study the role of the number ratio $N_b/N_f$.
We carried out a series of measurements at a fixed scattering length of $a_{bf}=1330a_0$ ($a_c/a_{bf}\approx 0.45$)~\cite{ac}. 
 As shown in Fig.~\ref{fig:vary_Q}, the largest frequency shift observed amounts to about 40\% for the smallest $N_b/N_f$, and it decreases to $\sim$10\% when $N_b/N_f$ increases from 0.03 to 0.19~\cite{SM}.

A theoretical description of the many-body dynamics of our Bose-Fermi system turns out to be rather challenging, because of the kinetics of the Fermi sea being essentially determined by the collisionless motion of the trapped fermions. The most simple model that captures the elementary features is an adiabatic Fermi sea (AFS) model, which assumes that the whole Fermi sea adapts adiabatically to the time-dependent mean-field formed by the BEC. This can be justified if at any position the local Fermi velocity is much larger than the speed of sound of the BEC~\cite{Yip2001cmi,Huang2018spa}. In addition we take advantage of the fermionic reservoir approximation~\cite{Lous2018pti}, which assumes a constant global chemical potential for the fermions. This leads to a time-dependent Gross-Pitaevskii equation for the BEC with a mean-field term from the fermions calculated in a quasi-stationary way. 
We solve the resulting differential equation numerically~\cite{SM} and show the results for the BEC breathing mode frequency by the dashed lines in Figs.~\ref{fig:vary_abf} and \ref{fig:vary_Q} for $a_c/a_{bf}>0.4$. For stronger repulsion strengths ($a_c/a_{bf}<0.4$), the extremely thin interface leads to numerical instabilities and challenges the basic assumption of an adiabatic fermion response.

Regarding the dependence on the strength of the repulsive effect (Fig.~\ref{fig:vary_abf}), the AFS model predictions agree with the measured points if the Fermi sea is not completely depleted in the trap center ($a_{bf} < a_d$). Beyond that, in the intermediate regime ($a_d \lesssim a_{\bf} \lesssim a_c$)  the model reproduces the emergence of a strong frequency up-shift, and it finally also shows the tendency of leveling-off in the limit of full phase separation (FPS), where $a_c/a_{bf} \rightarrow 0$. Qualitative agreement is also found in the dependence of the up-shift value on the number ratio (the dashed curve in Fig.~\ref{fig:vary_Q}) in the FPS regime.
Quantitatively, however, the frequency change calculated at $a_c/a_{bf}\approx 0.45$ is about 1.5 times larger than observed experimentally. 

For the case of full phase separation, we develop another approach, named the FPS model, to calculate the frequency shift ~\cite{Schaeybroeck2009tps, SM, Huang2018spa}.
Instead of assuming a quasi-static behavior of the Fermi sea, the FPS model describes the full dynamic response of a trapped Fermi sea. Intuitively, it embodies the trajectories of individual fermions, which repeatedly bounce off the interface and fall back to it at time intervals up to half of the fermion oscillation period $\pi/\omega_f$. Based on the collisionless Boltzmann transport equation, we calculate the dynamic response of the Fermi pressure at the oscillating Bose-Fermi interface. Then the frequency is obtained by matching the pressure and the radial speed at the interface. We find that the FPS model (solid curves in Fig.~\ref{fig:vary_abf} and \ref{fig:vary_Q}) gives a frequency shift very similar to the AFS results in the regime of full phase separation for all $N_b/N_f$ values that we have studied. Therefore we conclude that the dynamic character of the response does not provide an explaination for the deviation from the experiment~\cite{SM,Huang2018spa}.
 
A possible reason for the deviation may be due to the excitation scheme, which involves rapid switching of the interaction and thus creates additional fermion excitations~\cite{Nascimbene2009coo}, which remain unresolved in the oscillation signal~\cite{SM} and contaminate our signal.
Another reason may be a finite-temperature effect. The thermal bosonic component overlaps with the fermions and, at large interspecies scattering lengths, this forms a hydrodynamic shell around the BEC, which may affect the whole oscillation spectrum. Further investigations will be necessary to fully account for all mechanisms contributing to the large breathing mode frequency shift.

In general terms, our work shows how a small-sized BEC serves as a probe in a quantum fluid and provides information on both the interaction regime and the local properties of the environment. The latter can be described in terms of a decomposition into moments, which couple differently to various collective modes. The local pressure couples to the monopole (breathing) mode, the pressure gradient to dipole modes, and more complex anisotropies to higher-order modes. In our specific situation, the dominant effect results from the Fermi pressure acting on the breathing mode, whereas many scenarios can be envisioned where higher moments will strongly affect the collective mode spectrum. This can be the case in inhomogeneous systems, in more complex trapping potentials, in anisotropic environments realized in dipolar quantum fluids \cite{Baranov2008tpi, Lahaye2009tpo, Burdick2016lls, Kadau2016otr, Trautmann2018dqm}, or in spin-orbit coupled systems \cite{Galitski2013soc, Zhai2015dqg}.
 
Bose-Fermi mixtures with tunable interactions represent promising systems for the realization of novel fermionic superfluids based on boson-mediated pairing effects \cite{Bijlsma2000pei, Heiselberg2000ioi, Efremov2002pwc, Suzuki2008pws, Enss2009snp, Caracanhas2017fbm, Kinnunen2018ipw}, for both strongly attractive and repulsive interspecies interactions. The current experimental possibilities are enhanced by the increasing number of mixtures available in the laboratory; see e.g.\ Refs.~\cite{Trautmann2018dqm, Ravensbergen2018poa} for recent examples. In all candidate systems for boson-mediated fermion pairing, an issue of crucial importance is the competition between the formation of pairing phases and the onset of instabilities. Our current studies unveil the elementary dynamics in a strongly repulsive Bose-Fermi mixture and point to more general ways to extract information from the collective dynamics in regimes of particular interest.

\begin{acknowledgments}
We acknowledge valuable discussions with M. Baranov, D. Yang, R. van Bijnen, B. van Schaeybroeck, A. Lazarides and T. Maruyama on the theoretical models. We also thank A. Bergschneider and T. W. Grogan for comments on the manuscript. We acknowledge support by the Austrian Science Fund FWF within the Spezialforschungsbereich FoQuS (F4004-N23) and within the Doktoratskolleg ALM (W1259-N27).
\end{acknowledgments}

\bibliographystyle{apsrev4-1}

\begin{thebibliography}{67}
\bibitem [{ODT()}]{ODT2beams}%
  \BibitemOpen
  \href@noop {} {}\bibinfo {note} {The two trapping laser beams have a
  wavelength of 1064~nm and are crossed at an angle of 16 degrees. The waist of
  the lasers are 44 and 60 $\mu$m, respectively.}\BibitemShut {Stop}%
\bibitem [{abb()}]{abb}%
  \BibitemOpen
  \href@noop {} {}\bibinfo {note} {The values for $a_{\rm bb}$ and $a_{\rm bg}$
  are accidentally the same within three digits.}\BibitemShut {Stop}%
\bibitem [{dam()}]{damping}%
  \BibitemOpen
  \href@noop {} {}\bibinfo {note} {The observed damping rate $\gamma/\omega_0$
  of the BEC breathing mode remains near 0.02 in the absence of the fermions.
  We attribute this to the properties of the trap and the bosons.}\BibitemShut
  {Stop}%
\bibitem [{ac()}]{ac}%
  \BibitemOpen
  \href@noop {} {}\bibinfo {note} {The ratio $a_c/a_{\rm bf}$ varies between
  0.43 and 0.47, because of small changes in $a_c$ with variations of $E_{\rm
  F}$ \cite{SM}}\BibitemShut {NoStop}%
\end{thebibliography}

\end{document}